# Investigation of Polarization-Dependent Optical Force in Optical Tweezers using Generalized Lorenz-Mie Theory

Jai-Min Choi

*Division of Science Education, Chonbuk National University, Jeonju 54896, Korea*

Heeso Noh

*Department of Physics, Kookmin University, Seoul 02707, Korea*

In vectorial diffraction theory, tight focusing of a linearly polarized laser beam produces an anisotropic field distribution around the focal plane. We present a numerical investigation of the electromagnetic field distribution of a focused beam in terms of the input beam polarization state and the associated effects on the trap stiffness asymmetry of optical tweezers. We also explore the symmetry change of a polarization-dependent optical force due to the electromagnetic field redistribution by the presence of dielectric spheres of selected diameters ranging from the Rayleigh scattering regime to the Mie scattering regime.

Email: heesonoh@kookmin.ac.kr

Fax: +82-2-910-4728

## I. INTRODUCTION

Achieving the tightest spot of a highly focused light requires a dedicated state-of-the-art optical system and a rigorous description of light-matter interaction, from which other tuning parameters for controlling optical field distribution might be determined. To realize the stable trap condition in optical tweezers (OT), a laser beam should be focused enough to overcome the scattering force using a high-numerical aperture (NA) objective lens [1]. Recent studies showed that spherical aberration (SA) exhibited by high-NA optical systems is due to a refractive index mismatch between media interfaces [2], and could be tailored using control parameters, e.g., the adjusted wavefront of an input laser beam [3] or the immersion medium refractive index [4]. The electromagnetic (EM) field distribution of a focused laser beam is anisotropic in the focal plane even in an aberration-free ideal optical system with rotational symmetry [5-10]. Anisotropic distribution of the focused EM field manifests itself as trap stiffness (spring constant) asymmetry in OTs that utilize a linearly polarized laser beam [8-10]. In this study, we address the EM field distribution anisotropy of a highly focused light in terms of the input beam polarization state, demonstrating that the associated stiffness asymmetry in optical tweezers could also be tuned via the trap laser beam polarization state.

For a precise description of the highly focused EM field, the vectorial nature of light needs to be considered, as formulated in the angular spectrum method [5, 11], whereas scalar diffraction theory predicts an Airy disk pattern for a typical optical system. The tightly focused EM field distribution of a linearly polarized input beam has been investigated in various situations; a homogeneous medium [5], a single interface dividing immersion and specimen media [11], and a more realistic situation consisting of three interfaces including coverslip thickness [2]. The focused EM field distribution of circularly polarized light has also been investigated with respect to angular momentum transfer using multipole expansion [12]. Assuming aberration free conditions (this could be achieved to some extent using a correction collar of a modern objective lens [3]), we investigated focused EM field anisotropy using vectorial diffraction theory. Polarization-dependent EM field distributions in highly focused OTs were evaluated numerically in terms of trap stiffness asymmetry, which could be interpreted as the landscape of optical potential. For this purpose, we revised the generalized Lorenz–Mie theory (GLMT) developed by Barton *et al.* [13-15] to include trap beam polarization state in optical force calculations.

## II. Mathematical Background for Numerical Calculations

In this section, we illustrate the polarization-dependent electric field distribution of a highly focused beam in a homogeneous medium using the angular spectrum method. We outline the procedure for a GLMT calculation of optical force that was revised to encompass the general polarization state of an input beam. The coordinate systems are depicted in Fig. 1. Cartesian coordinate systems $(x,y,z)$ and $(x',y',z')$ are centered on the lens aperture $\Sigma$ and the focal point $O_G$, respectively, of the focused beam, with the same orientations. The other, unannotated, Cartesian system centered at $O_P$ is particle-centered for the GLMT calculation. The collimated input beam represented by its electric field $\vec{E}^{in}$ enters the lens aperture $\Sigma$ along the direction defined by wave vector $\vec{k}_0$, parallel to the optic axis ($z$ and $z'$), and is focused by the high-NA optical system into the medium space with a refractive index of $n_m$. One partial plane wave component, $\vec{E}^m$, in direction $\vec{k}$ is presented in Fig. 1, with a polar angle $\theta$. The other symbols for angles in Fig. 1 are implicit. The resulting electric field distribution at $\vec{r}_P(r_P,\theta_P,\phi_P)$ around the focal point ($O_G$) is represented by the angular spectrum method, which sums up all relevant partial wave components, as follows [2, 5, 10, 11]:

$$\vec{E}(r_P,\theta_P,\phi_P) = -\frac{ikf}{2\pi}\sqrt{\frac{n_m}{n_0}} \int_0^{\alpha_{max}} d\theta \int_0^{2\pi} d\phi \times J(\theta)A(\theta)e^{-\sin^2\theta/(f_0^2 \sin^2\alpha_{max})} \mathbf{M}\vec{E}^{in} e^{ikr_P\kappa(\theta_P,\phi_P,\theta,\phi)}, \quad (1)$$

where $k$ is the wavenumber in the medium ($k = n_m k_0$), $J(\theta) = \sin\theta\cos\theta$ is the Jacobian for transformation from $(\hat{k}_x,\hat{k}_y)$ to $(\theta,\phi)$, the apodization factor $A(\theta) = 1/\sqrt{\cos\theta}$ fulfilling the sine condition, $f_0 = w_0/(f\sin\alpha_{max})$ is the fill factor for collimated input beam spot size $w_0$, $f$ is the focal length, the maximum polar angle is given as $\alpha_{max} = \mathrm{asin}(NA/n_m)$, $\mathbf{M}$ is the polarization matrix,

$\kappa(\theta,\phi,\theta_P,\phi_P) = \sin\theta\sin\theta_P\cos(\phi-\phi_P) + \cos\theta\cos\theta_P$, and integration is performed over the solid angle defined by $0 \leq \phi < 2\pi$ and $0 \leq \theta \leq \alpha_{max}$. The polarization matrix $\mathbf{M}$ given in Eq. (2) consists of the rotation matrix $\mathbf{R}$ about the optic axis and the matrix of lens action $\mathbf{L}$:

$$\mathbf{M} = \mathbf{R}^{-1}\mathbf{L}\mathbf{R} = \begin{pmatrix} \cos\phi & -\sin\phi & 0 \\ \sin\phi & \cos\phi & 0 \\ 0 & 0 & 1 \end{pmatrix} \begin{pmatrix} \cos\theta & 0 & \sin\theta \\ 0 & 1 & 0 \\ -\sin\theta & 0 & \cos\theta \end{pmatrix} \begin{pmatrix} \cos\phi & \sin\phi & 0 \\ -\sin\phi & \cos\phi & 0 \\ 0 & 0 & 1 \end{pmatrix}. \quad (2)$$

For a more practical description that includes media interfaces, a matrix $\mathbf{I}$ consisting of the Fresnel coefficients for *s*- and *p*-polarization components and changes to polar angle $\theta$ due to refraction need to be incorporated into $\mathbf{M}$ and $\kappa$ [2, 11]. The incident electric field is prepared for a general polarization state: $\vec{E}^{in} = (E_x, E_y, 0)$, where $E_x$ and $E_y$ are complex numbers; a numerical evaluation of Eq. (1) then provides the polarization-dependent field distribution around the focal plane. The focused magnetic field distribution could be obtained using a similar procedure.

Highly focused electric field patterns are obtained through numerical integration of Eq. (1) for two extreme polarization states: the linear and circular polarization states. We consider the following parameters: NA = 1.2, $f$ = 3 mm, input beam vacuum wavelength $\lambda_0$ = 780 nm, $f_0$ = 1, and medium refractive index $n_m$ = 1.33. Electric field components $(|E_{x'}|, |E_{y'}|, |E_{z'}|)$ and normalized intensity $\tilde{I}$ in the focal plane ($x'y'$-plane) are presented in Fig. 2(a) for a linearly polarized incident field: $\vec{E}^{in} = E_0\hat{x}$, where the electric field components and intensity are normalized such that $|E_{x'}| = 1$ at the focal point $O_G$. The frame ticks in Fig. 1 are given in units of the minimum spot size $w_G$ = 293 nm, obtained from a Gaussian fit of the numerically calculated field distribution. The white lines (dashed line) outline the $5\times 10^{-2}$ level of the magnitude of the electric field components (the $e^{-2}$ level of $\tilde{I}$). Additional polarization components, $E_y$ and $E_z$, produce the anisotropic field distribution, thus resulting in an elongated intensity distribution. The main contribution to the anisotropic intensity distribution originates from the *z*-polarization component: $(|E_{x'}|, |E_{y'}|, |E_{z'}|)_{\max}^{\text{LP}} = (1, 0.06, 0.34)$. The aspect ratio of the elongated intensity distribution is $(d_{\hat{y}'}^{\text{LP}}/d_{\hat{x}'}^{\text{LP}} =)$ 0.79: $d_{\hat{x}'}^{\text{LP}} = 1.22 w_G$ and $d_{\hat{y}'}^{\text{LP}} = 0.96 w_G$ (see Fig. 2). Compared to the anisotropic field distribution of the linearly polarized input beam, a circularly polarized input beam renders an isotropic field distribution in the focal plane. In Fig. 2(b), electric field components $(|E_{x'}|, |E_{y'}|, |E_{z'}|)$ and the normalized intensity $\tilde{I}$ in the focal plane are presented for a circularly polarized incident field: $\vec{E}^{in} = E_0(\hat{x} + i\hat{y})/\sqrt{2}$. The aspect ratio of the intensity distribution resulting from circularly polarized light is 1 within the precision of numerical calculation, i.e., $d_{\hat{x}'}^{\text{CP}} = d_{\hat{y}'}^{\text{CP}} = 1.12\ w_G$, and the maximum electric field strengths of the polarization components $(|E_{x'}|, |E_{y'}|, |E_{z'}|)_{\max}^{\text{CP}} = (0.71, 0.71, 0.24)$. This results are confirmed with the numerical simulation using finite difference time domain (FDTD) method (not shown).

For small particles ($\ll \lambda$), Rayleigh approximation is valid and a dielectric sphere can be treated as a point dipole with dipole moment $\vec{p} = 4\pi\varepsilon_p a^3 ((\bar{n}^2 - 1)/(\bar{n}^2 + 2))\vec{E}$, where $\varepsilon_p$ is the permittivity of a particle, $a$ is the particle radius, the relative refractive index $\bar{n}$ is the ratio of the refractive index of particle $n_p$ to that of medium $n_m$, and $\vec{E}$ is the electric field. In the Rayleigh scattering regime, exerted radiation force can be represented in terms of the gradient force, $\vec{F}_{grad} = \left\langle \left[ \vec{p}(\vec{r}',t) \cdot \vec{\nabla} \right] \vec{E}(\vec{r}',t) \right\rangle_T$, and the scattering force, $\vec{F}_{sc} = \frac{n_m}{c} C_{sc} I(\vec{r}')\hat{z}'$, where $c$ is the speed of light, $C_{sc}$ is the scattering cross section, $I(\vec{r}')$ is the laser beam intensity, and $\langle \ \rangle_T$ denotes time average for reaching a steady state [16]. However, when particle size becomes comparable to the wavelength of the trap laser beam, the EM field is not constant over particle and the presence of the particle modifies the EM field distribution. A rigorous GLMT method is required to calculate optical force in the Mie scattering regime ($2\pi a \geq \lambda$).

Optical force is given by the surface integral of Maxwell's stress tensor, the integrand in Eq. (3), over the enclosing surface $S$ of a spherical particle (see Fig. 1) [13, 15]:

$$\langle \vec{F} \rangle_T = \int_S \left\langle \varepsilon E_r \vec{E} + \mu H_r \vec{H} - \frac{1}{2}(\varepsilon E^2 + \mu H^2)\hat{r} \right\rangle_T da , \qquad (3)$$

where the associated EM fields, $\vec{E} = \vec{E}^{in} + \vec{E}^{sc}$ and $\vec{H} = \vec{H}^{in} + \vec{H}^{sc}$, in the Maxwell stress tensor consist of the incident field, $\vec{E}^{in}$ and $\vec{H}^{in}$, and the scattered field, $\vec{E}^{sc}$ and $\vec{H}^{sc}$, from a particle in the surrounding medium, $\varepsilon$ and $\mu$ are the permittivity and permeability of the surrounding medium, respectively, and $\hat{r}$ is the unit vector along the radial direction ($\vec{r}_a$) in spherical coordinates, with the origin at $O_p$. For a spherical particle of radius $a$, EM fields can be represented by a series expansion of the Riccati–Bessel functions [13, 14], $\psi_l(r)$ and $\xi_l(r)$, and spherical harmonics $Y_{lm}(\theta'',\phi'')$ with the following expansion coefficients: ($A_{lm}$, $B_{lm}$) for the incident field and ($a_{lm}$, $b_{lm}$) for the scattered field, and ($l$, $m$) are the spherical harmonic mode numbers.. The boundary condition across the sphere surface provides the following relations between ($a_{lm}$, $b_{lm}$) and ($A_{lm}$, $B_{lm}$) [13]:

$$a_{lm} = \frac{\psi_l'(\bar{n}\alpha)\psi_l(\alpha) - \bar{n}\psi_l(\bar{n}\alpha)\psi_l'(\alpha)}{\bar{n}\psi_l(\bar{n}\alpha)\xi_l'(\alpha) - \psi_l'(\bar{n}\alpha)\xi_l(\alpha)} A_{lm},$$

$$b_{lm} = \frac{\bar{n}\psi_l'(\bar{n}\alpha)\psi_l(\alpha) - \psi_l(\bar{n}\alpha)\psi_l'(\alpha)}{\psi_l(\bar{n}\alpha)\xi_l'(\alpha) - \bar{n}\psi_l'(\bar{n}\alpha)\xi_l(\alpha)} B_{lm} \qquad (4)$$

, where $\alpha = n_m k_0 a$ and the apostrophe (`) denotes the radial derivative. The highly focused EM field can also be represented in a functional form using the fifth-order Gaussian beam method [14], where the characteristic beam spot size $w_G$ needs to be plugged in considering the parameters of a virtual optical tweezers setup. This enables the extraction of coefficients $A_{lm}$ and $B_{lm}$ via numerical integration of the EM field function multiplied by $Y_{lm}^*(\theta'',\phi'')$, where "*" denotes complex conjugation. Knowledge of the incident and scattered EM field coefficients, together with an appropriately chosen truncation mode number $l = l_{max}$ allows the evaluation of optical force given in Eq. (3). To incorporate a general polarization state in the GLMT calculation, EM field components of the fifth-order Gaussian beam [14] are transformed using standard Jones matrices for rotation and phase retardation.

### III. Results and Discussion

Numerical investigation of the polarization dependence of optical force in OTs begins with a linearly polarized trap beam and a homogeneous polystyrene (PS) sphere as the trap object. The following parameters were used for virtual OT experiments: $\lambda_0$ = 780 nm, NA = 1.2 (water immersion), $f$ = 3 mm, $f_0$ = 1, $n_m$ = 1.33, $n_p$ = 1.57, and beam power $P_0$ in the trap region is 10 mW. Radiation forces in transverse and longitudinal directions are presented in Fig. 3(a-c) for particles with radii $a$ = (50, 150, 500) nm, respectively. The transverse force components, $F_{\hat{x}'}$ (red solid line) and $F_{\hat{y}'}$ (blue dashed line), were estimated at equilibrium positions $z'_{eq}$ = (20, 205, 226) nm for $a$ = (50, 150, 500) nm, respectively, where the longitudinal force component, $F_{\hat{z}'}$ (green solid line), is zero; the scattering force pushed the equilibrium position away from the focal point. The trap potential anisotropy in the transverse plane was estimated in terms of trap stiffness asymmetry factor $s_T = 1 - k_{\hat{x}'}/k_{\hat{y}'}$ where trap stiffness was deduced using the nonlinear fit of the numerically calculated optical force for the corresponding degree of freedom. For the 100 nm PS spheres (Fig. 3(a)) the trap stiffnesses ($k_{\hat{x}'}, k_{\hat{y}'}$) were (0.65, 1.16) pN/μm, yielding a transverse plane stiffness asymmetry factor of $s_T$ = 0.44. For the 300 nm PS sphere (Fig. 3(b)), the trap stiffnesses and stiffness asymmetry factor were (13.5, 19.3) pN/μm and $s_T$ = 0.3, respectively. Results for the 1 μm PS sphere (Fig. 3(c)) clearly indicate not only an averaging effect on the EM field distribution over the particle, but also EM field distribution modification due to particle presence: the trap stiffnesses ($k_{\hat{x}'}, k_{\hat{y}'}$) of the 1 μm PS sphere were (33.3, 29.8) pN/μm, with a corresponding stiffness asymmetry factor $s_T$ = -0.12 < 0. This counterintuitive sign change of the asymmetry factor was reported in previous studies using a linearly polarized trap beam [8-10].

The GLMT-based optical force calculations in transverse directions are presented in Fig. 4 for a circularly polarized trap beam and PS spheres of radius (a) 100 nm, (b) 300 nm, and (c) 1 µm. Compared to the linear polarization results, orthogonal pairs of lateral forces overlapped within the precision of numerical calculation for the given particle sizes, implying that the isotropic field distribution provided by circularly polarized light is preserved even in the presence of a trapped spherical particle. The trap stiffnesses and asymmetry factors for circular polarization are summarized in Table 1, together with the linear polarization results. We also present the longitudinal stiffness asymmetry factor, $s_L = (k_{\hat{x}'} + k_{\hat{y}'})/2k_{\hat{z}'}$, which is the ratio of the mean lateral stiffness to the axial stiffness. The numerical values of $s_T^{CP}$ only represent the numerical error; the symmetry of EM field distribution and the test spheres ensures $s_T^{CP}$ to be 0 (zero).

We now apply our numerical method to Rohrbach's experimental results [8], where the trap stiffness asymmetry of highly focused OTs formed by focusing linearly polarized light was investigated experimentally and theoretically using a two-component model [8, 17]. The experimental parameters were as follows; $\lambda_0$ = 1064 nm, $P_0$ = 10 mW, $n_m$ = 1.33, $n_p$ = 1.57, NA =1.2, $f_0$ = 2, and particle radii of (110, 265, 345, 425, 515, and 830) nm. Using these parameters, the minimum beam spot size $w_G$ was estimated to be 399 nm; we assumed an aberration-free ideal optical system. In Fig. 5, filled and empty black circles denote experimental and theoretical values from Rohrbach's study [8], respectively. Experimental data indicated that the trap stiffness asymmetry factors for PS particles larger than the beam wavelength ($\lambda_0/n_m$) were negative; $k_x < k_y$ for particle radii $a < \lambda_0/n_m$; and $k_x > k_y$ for particle radii $a > \lambda_0/n_m$. Rohrbach's theoretical estimation of trap anisotropy using the two-component model developed in the Rayleigh-Gans regime did not accurately describe the experimental data for particle sizes comparable to the beam wavelength. Significant gaps between experimental and theoretical results existed except the two smallest particles ($a$ = 110, 265 nm); their theoretical method could not support the experimentally observed trap asymmetry factor sign change. Our calculation results based on revised GLMT show good agreement with Rohrbach`s experimental data, including the trap anisotropy sign change. More importantly, a demonstration of trap anisotropy control using the polarization state of the trap beam is presented (blue diamonds in Fig. 5). With a circularly polarized trap beam, trap stiffness asymmetry factors were numerically estimated to be two orders of magnitude less than the linear polarization case for all particle sizes. We note that the trap asymmetry factor should be zero in an ideal case, because of an isotropic field distribution of circularly polarized light in the transverse plane and homogeneous particles of spherical shape.

## IV. Conclusion

We investigated polarization-dependent optical forces using a revised GLMT method that incorporated the general polarization state and minimum beam spot size. Dependence of trap anisotropy on the particle size was investigated for a linearly polarized trap beam and compared with available experimental data, providing a better agreement than models employed in a previous study [8], especially for large spheres. Our results indicate that the trap stiffness asymmetry factor strongly depends on the particle size and the input beam polarization state. The trap anisotropy modified by the presence of a particle can be readjusted via trap beam polarization control to be isotropic. We believe that our investigation of polarization-dependent optical forces may be useful for precise control of particle dynamics in colloidal systems and in realizing more ideal systems for parametric resonance studies in the context of OTs.

## ACKNOWLEDGEMENTS

This work was supported by the Chonbuk National University research fund in 2014 and the Basic Science Research Program through the National Research Foundation of Korea (NRF) funded by the Ministry of Education, Science, and Technology (No. 2011-0014908).

**Table 1.** Trap stiffness and corresponding asymmetry factor in the transverse plane for various PS sphere radii, and input beam polarization state. The superscript distinguishes polarization states: LP and CP for linear and circular polarization state, respectively. Parameters for this calculations are as follows: $\lambda_0$ = 780 nm, NA = 1.2, $f$ = 3 mm, $f_0$ = 1, $n_m$ = 1.33, $n_p$ = 1.57, $\lambda_n = \lambda_0 / n_m$, $\Delta\phi = (\bar{n}-1)k2a$, and $P_0$ = 10 mW. The unit of stiffness is pN/$\mu$m.

| Radius (nm) | $\Delta\phi$ | $k_{\hat{z}'}$ | $(k_{\hat{x}'}, k_{\hat{y}'})^{LP}$ | $(s_T, s_L)^{LP}$ | $(k_{\hat{x}'}, k_{\hat{y}'})^{CP}$ | $(s_T, s_L)^{CP}$ |
|---|---|---|---|---|---|---|
| 50 | 0.19 | 0.27 | (0.65, 1.16) | (0.44, 3.4) | (0.90, 0.90) | (2.0×10$^{-3}$, 3.3) |
| 150 | 0.58 | 2.8 | (13.5, 19.3) | (0.30, 5.9) | (16.5, 16.6) | (6.1×10$^{-3}$, 5.9) |
| 250 | 0.97 | 5.5 | (39.2, 42.5) | (0.08, 7.4) | (40.7, 40.8) | (1.4×10$^{-3}$, 7.4) |
| 350 | 1.35 | 8.7 | (53.3, 46.8) | (-0.14, 5.8) | (50.1, 50.0) | (-1.9×10$^{-3}$, 5.8) |
| 500 | 1.93 | 7.3 | (33.3, 29.8) | (-0.12, 4.3) | (31.4, 31.5) | (-3.2×10$^{-3}$, 4.3) |

**Figure Captions**

**Fig. 1**. Schematic diagram of coordinate systems used for focused field and optical force calculations based on revised GLMT.

**Fig. 2**. Normalized electric field components $\left(|E_{\hat{x}'}|, |E_{\hat{y}'}|, |E_{\hat{z}'}|\right)$ and normalized intensity distribution $\tilde{I}$ at the focal plane ($x'y'$-plane). Frame ticks are given in units of the minimum spot size, $w_G$. (a) Linearly and (b) circularly polarized trap beams.

**Fig. 3**. GLMT calculation of radiation force for linearly polarized light. PS sphere radii: (a) 50 nm, (b) 150 nm, and (c) 500 nm. For each PS sphere radius (a-c), transverse force components $F_{\hat{x}'}$ (red solid line) and $F_{\hat{y}'}$ (blue dotted line) are presented in the left panel and longitudinal force $F_{\hat{z}'}$ (green solid line) is presented in the right panel.

**Fig. 4**. GLMT calculation of radiation forces for circularly polarized light. PS sphere radii: (a) 50 nm, (b) 150 nm, and (c) 500 nm. For each PS sphere radius (a-c), transverse force components, $F_{\hat{x}'}$ and $F_{\hat{y}'}$, are denoted by red and blue dashed lines, respectively.

**Fig. 5**. Stiffness asymmetry versus particle size, depending on polarization state. Filled and empty circles denote Rohrbach's experimental data and theoretical values [8], respectively. Red empty squares and blue diamonds denote our theoretical estimations for linear polarization (LP) and circular polarization (CP) states, respectively.

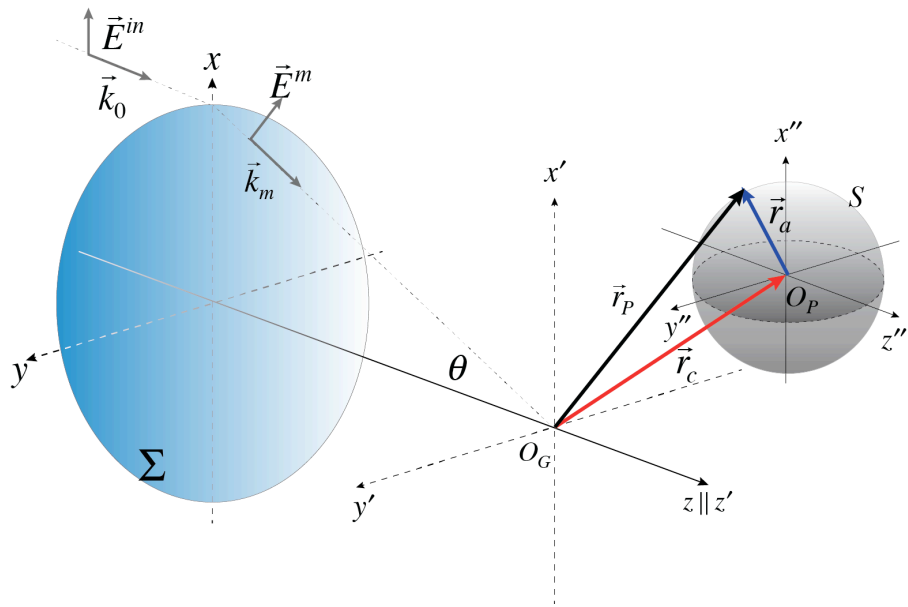

Fig. 1

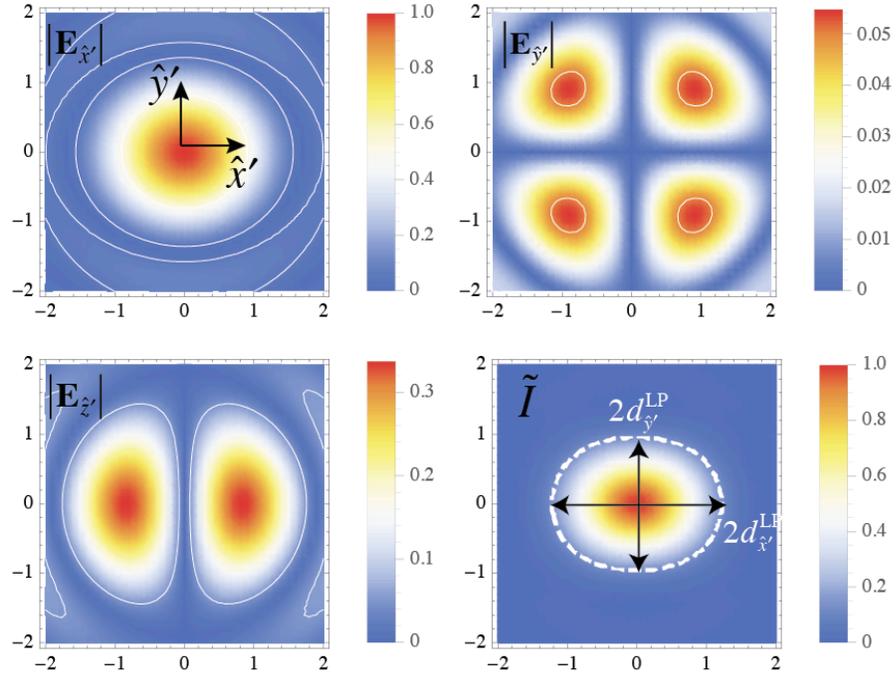

(a)

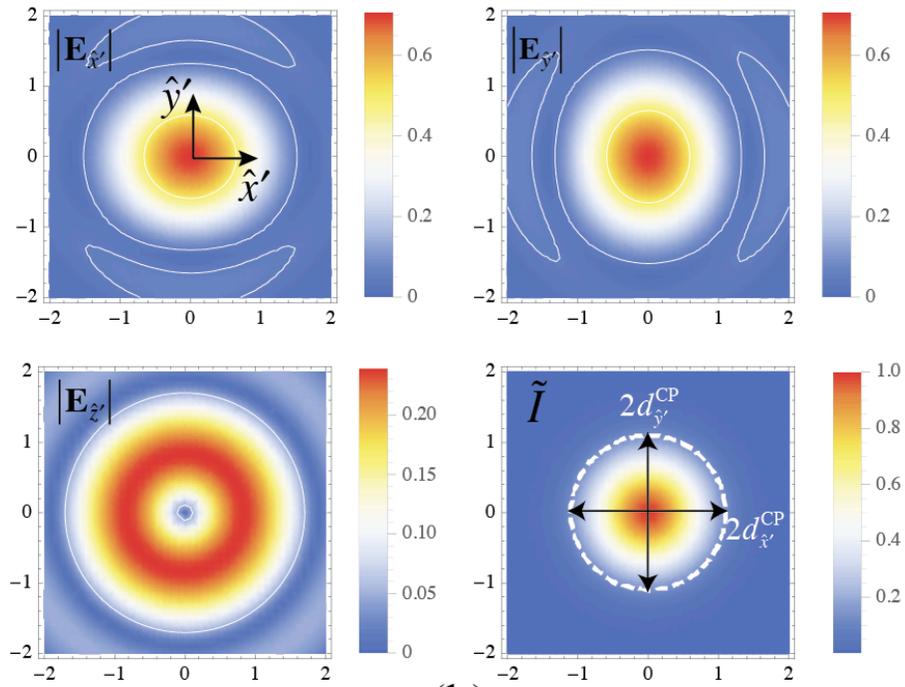

(b)

Fig. 2

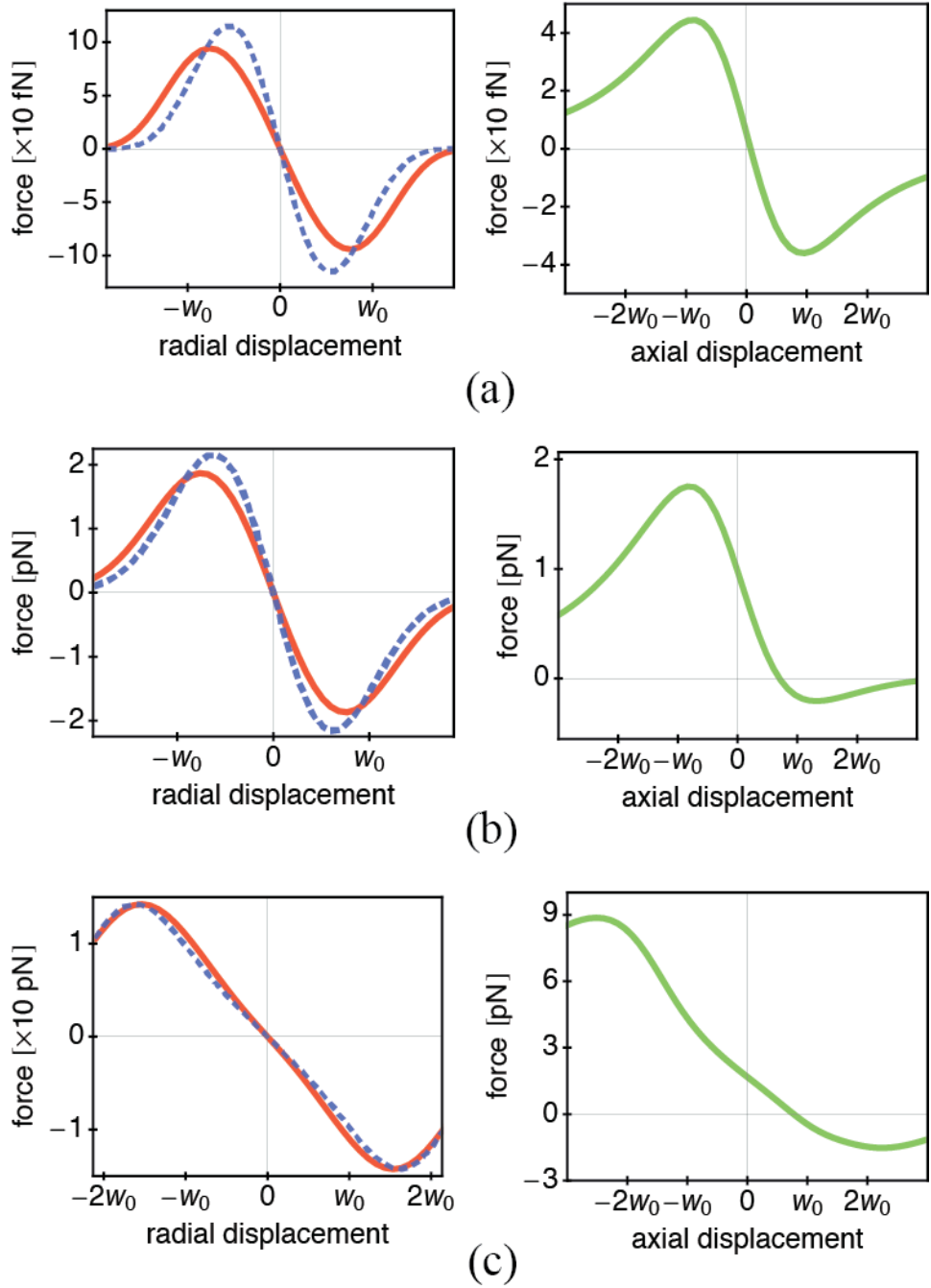

Fig. 3

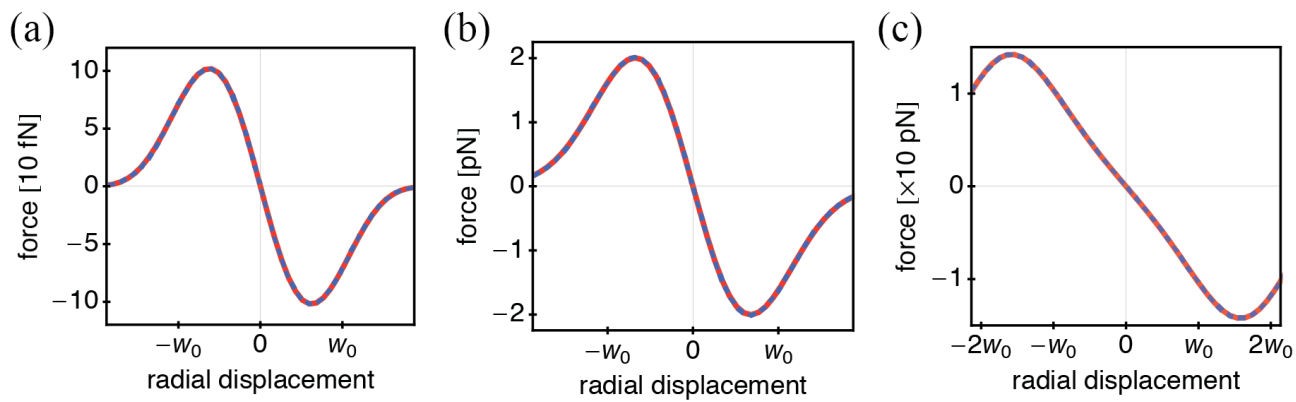

Fig. 4

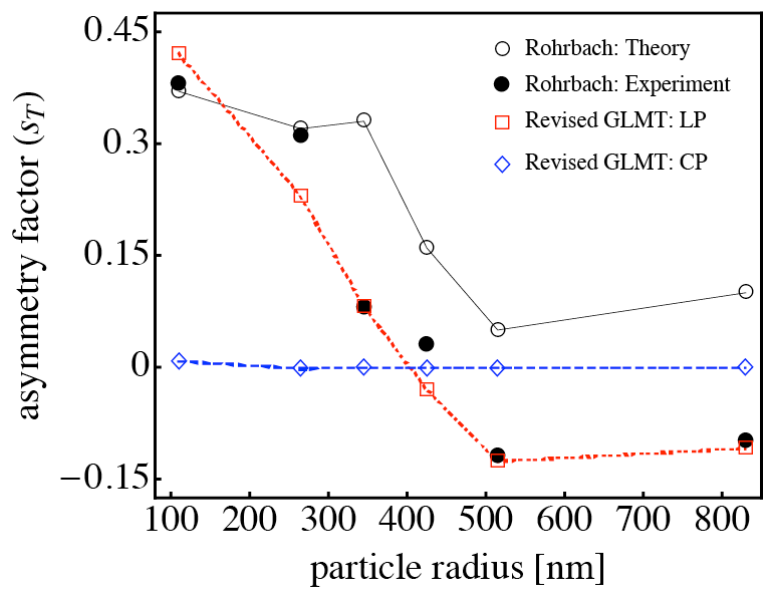

Fig. 5